\documentclass[12pt,a4paper,twoside]{article}
\RequirePackage{latexsym,amsmath,amssymb}
\RequirePackage[dvipsnames,usenames]{color}
\usepackage{a4wide}
\usepackage{amssymb}
\usepackage{bm}
\usepackage{dcolumn}
\usepackage{amsfonts}
\usepackage{graphicx,epsfig}
\usepackage{psfrag}
\usepackage{multicol}
\usepackage{tikz}
\usetikzlibrary{arrows,shapes}
\usetikzlibrary{matrix,arrows}
\newcommand{\be}{\begin{eqnarray}}
\newcommand{\ee}{\end{eqnarray}}

\newcommand{\ket}[1]{\mbox{$\mid #1\,\rangle$}}

\newcommand{\pro}[2]{\mbox{$\langle\, #1 \mid #2\,\rangle$}}
\newcommand{\expec}[1]{\mbox{$\langle\, #1\,\rangle$}}

\renewcommand{\d}{\mbox{${\rm d}$}} 
\newcommand{\lp}{\ell_{\rm p}}
\newcommand{\mpl}{m_{\rm p}}

\newcommand{\rh}{r_{\rm H}}
\newcommand{\Rh}{R_{\rm H}}
\newcommand{\psis}{{\psi}_{\rm S}}

\newcommand{\psih}{{\psi}_{\rm H}}
\title{\bf Horizons and non-local time evolution of quantum mechanical systems}
\author{Roberto~Casadio\thanks{E-mail: casadio@bo.infn.it}
\\
\\
{\em $^a$Dipartimento di Fisica e Astronomia, Universit\`a di Bologna}
\\
{\em via Irnerio~46, I-40126 Bologna, Italy}
\\
\\
{\em $^b$I.N.F.N., Sezione di Bologna,}
\\
{\em via B.~Pichat~6/2, I-40127 Bologna, Italy}
}
\begin{document}
\maketitle
\begin{abstract}
According to general relativity, trapping surfaces and horizons are
classical causal structures that arise in systems with sharply defined
energy and corresponding gravitational radius.
The latter concept can be extended to a quantum mechanical matter state
simply by means of the spectral decomposition, which allows one to define an
associated ``horizon wave-function''.
Since this auxiliary wave-function contains crucial information about the causal
structure of space-time, a new proposal is formulated for the time-evolution
of quantum systems in order to account for the fundamental classical property
that outer observers cannot receive signals from inside a horizon.
The simple case of a massive free particle at rest is used throughout
the paper as a toy model to illustrate the main ideas.
\end{abstract}
\section{Introduction}
\label{intro}
\setcounter{equation}{0}
Non-locality in modern theoretical physics is certainly a key issue, as there are many
hints suggesting that the evolution of quantum mechanical states cannot be properly
described without taking into account the extension (in space and time) of the whole
physical system under consideration (or, naively, its whole wave-function in position
space).
Non-locality also appears at the classical level in systems in which gravity plays a
crucial role, a remarkable example being given by trapping surfaces and horizons.
In fact, the location of a trapping surface can be determined by just considering the
metric locally, however the metric itself at a point will in general depend on all of the
matter sources around, and in a complicated non-linear manner.
An event horizon can then only be identified provided one knows the entire history
of the space-time.
In any quantum theory of gravity, it is therefore likely that space and time non-locality
will become one of the most prominent features overall.
\par
Unusual causal structures like the trapping surfaces and horizons, could only occur
in strongly gravitating systems, such as astrophysical objects that collapse and
possibly form black holes.
One might argue that, for a large black hole, gravity should appear ``locally weak'' at
the horizon, since tidal forces look small to a freely falling observer (their magnitude
being roughly controlled by the surface gravity which is inversely proportional to
the horizon radius).
However, light (like any other classical signal) is confined inside the horizon,
no matter how weak such local forces may appear to a local observer, which
we could take as the definition of a ``globally strong'' interaction.
Moreover, for a small black hole (with a mass about the Planck scale),
tidal forces become strong both in the local and global sense, thus granting
such an energy scale a remarkable role in the search for a quantum theory
of gravity.
It is indeed not surprising that modifications to the standard commutators of
quantum mechanics and Generalised Uncertainty Principles (GUPs) have
been proposed, essentially in order to account for the possible existence of small
black holes around the Planck scale, and the ensuing minimum measurable
length~\cite{hossenfelder}.
Unfortunately, that regime is presently well beyond our experimental capabilities,
at least if one takes the Planck scale at face value.
Nonetheless, there is the possibility that the low energy theory still retains
some signature features that could be accessed in the near future (see, for 
example, Refs.~\cite{hogan}).
\par
Before we start calculating phenomenological predictions, it is of the foremost
importance that we clarify the possible conceptual issues arising from the use of
arguments and observables that we know work at our every-day scales.
One of such key concepts is the gravitational radius of a self-gravitating
source, which can be used in order to asses the existence of trapping
surfaces, at least in spherically symmetric systems, whose metric
$g_{\mu\nu}$ can be written as~\footnote{We shall use units with $c=1$,
and the Newton constant $G=\lp/\mpl$, where $\lp$ and $\mpl$
are the Planck length and mass, respectively, and $\hbar=\lp\,\mpl$.}
\be
\d s^2
=
g_{ij}\,\d x^i\,\d x^j
+
r^2(x^i)\left(\d\theta^2+\sin^2\theta\,\d\phi^2\right)
\ ,
\label{metric}
\ee
where $r$ is the areal coordinate and $x^i=(x^1,x^2)$ are coordinates
on surfaces of constant angles $\theta$ and $\phi$.
The location of a trapping surface is then determined by the equation
\be
g^{ij}\,\nabla_i r\,\nabla_j r
=
0
\ ,
\label{th}
\ee
where $\nabla_i r$ is perpendicular to surfaces of constant area
$\mathcal{A}=4\,\pi\,r^2$.
If we set $x^1=t$ and $x^2=r$, and denote the matter density as
$\rho=\rho(r,t)$, Einstein field equations tell us that
\be
g^{rr}
=
1-\frac{2\,\lp\,(m/\mpl)}{r}
\ ,
\label{einstein}
\ee
where the Misner-Sharp mass is given by 
\be
m(r,t)
=
4\,\pi\int_0^r \rho(\bar r,t)\,\bar r^2\,\d \bar r
\ ,
\label{M}
\ee
as if the space inside the sphere were flat.
A trapping surface then exists if there are values of $r$ (and $t$) such that
the gravitational radius
\be
\Rh
=
2\,\lp\,\frac{m}{\mpl}
\ ,
\label{hoop}
\ee
satisfies 
\be
\Rh(r,t)\ge r
\ .
\label{Ehor}
\ee
If the above relation holds in the vacuum outside the region where the source is located,
$\Rh$ becomes the usual Schwarzschild radius, and the above argument gives
a mathematical foundation to Thorne's {\em hoop conjecture\/}~\cite{Thorne:1972ji},
which (roughly) states that a black hole forms when the impact parameter $b$
of two colliding small objects is shorter than the Schwarzschild radius of the system,
that is for $b \lesssim 2\,\lp\,{E}/{\mpl}$, where $E$ is the total energy in the
centre-of-mass frame. 
\par
The above treatment becomes questionable for sources of the Planck size or
lighter, for which we know for an experimental fact that quantum effects
may not be neglected.
Consider in fact a spin-less point-like particle of mass $m$, whose Schwarzschild
radius is given by $\Rh$ in Eq.~\eqref{hoop} with $m$ now a constant.
The Heisenberg principle of quantum mechanics introduces an uncertainty in
the particle's spatial localisation of the order of the Compton-de~Broglie length,
$\lambda_m \simeq \lp\,{\mpl}/{m}$.
Since quantum physics is a more refined description of reality, we could argue that
$\Rh$ only makes sense if~\footnote{Quite notably, this argument also explains
why one expects quantum gravity to become relevant at the Planck mass, which
could otherwise be just a numerological accident.} 
\be
\Rh\gtrsim \lambda_m
\quad
\Rightarrow
\quad
m
\gtrsim
\mpl
\ ,
\label{clM}
\ee
which brings us to face a conceptual challenge:
how can we describe quantum mechanical systems (like the elementary particles)
which are classically expected to have a horizon smaller than the size of the
uncertainty in their position?
\par
In Refs.~\cite{fuzzyh}, a proposal was put forward in order to describe the
``fuzzy'' Schwarzschild (or gravitational) radius of a localised (but likewise
fuzzy) quantum source, which was then shown to induce a GUP and minimum
measurable length~\cite{gupf}, and also yields corrections to the classical hoop
conjecture~\cite{acmo}.
The same approach shows that Bose-Einstein condensate (BEC) models
of black holes~\cite{dvali} actually possess a horizon, with a proper
semiclassical limit~\cite{BEC_BH}.
It is important to emphasise that our approach differs from most previous attempts
in which the gravitational degrees of freedom of the horizon, or of the black hole metric,
are quantised independently of the nature and state of the source (for some
bibliography, see, e.g., Ref.~\cite{davidson}). 
In our case, the gravitational radius is instead quantised along with the matter
source that produces it, somewhat more in line with the highly non-linear general
relativistic description of the gravitational interaction.
However, having given a practical tool for describing the gravitational radius
of a generic quantum system is just the starting point.
In fact, when the probability that the source is localised within its gravitational
radius is significant, the system should show (some of) the properties ascribed 
to a black hole in general relativity.
These properties, the fact in particular that no signal can escape from the interior,
only become relevant once we consider how the overall system evolves. 
In this work, we shall hence address the crucial issue of the time evolution
of quantum states with a ``fuzzy'' horizon. 
\par
In the next Section, we shall first review the general idea of the horizon
wave-function, and then analyse in detail the case of a massive spherical
particle at rest.
In Section~\ref{Tevo}, the spherical particle will serve as a toy model
to investigate a proposal for a modified, causal evolution, in which the possible
presence of a horizon is taken into account.
We shall finally comment on our results and some of the limitations of our
approach in Section~\ref{conc}.
\section{Static horizon wave-function}
\label{PsiH}
\setcounter{equation}{0}
Let us start by reviewing the wave-function for the gravitational radius that can
be associated with any localised quantum mechanical particle~\cite{fuzzyh}.
As we mentioned in the Introduction, this wave-function emerges from relating
the matter source to its gravitational radius at the quantum level, rather than
considering quantum gravitational degrees of freedom independently~\cite{davidson},
and allows us to put on more quantitative grounds the condition~\eqref{clM}
that should distinguish black holes from regular particles.
\subsection{Formal definitions}
\label{defH}
For the sake of clarity, we shall just consider quantum mechanical states
representing spherically symmetric sources which are both {\em localised in space\/}
and {\em at rest\/} in the chosen reference frame.
In fact, localisation is an essential ingredient for generating trapping surfaces and
black holes, and we wish to avoid the complications due to the relative motion of the
source and departure from sphericity.
We shall also ignore any possible time evolution for now.
Our ``particle-like'' state will consequently be described by a wave-function
$\psis\in L^2(\mathbb{R}^3)$, which can be decomposed into energy eigenstates,
\be
\ket{\psis}
=
\sum_E\,C(E)\,\ket{\psi_E}
\ ,
\ee
where the sum represents the spectral decomposition in Hamiltonian eigenmodes,
\be
\hat H\,\ket{\psi_E}=E\,\ket{\psi_E}
\ ,
\ee
and $H$ should be specified depending on the system we wish to consider.
\par
We can now try and quantise the gravitational radius determined by the
Misner-Sharp mass~\eqref{M}, by simply expressing the energy in terms
of the Schwarzschild radius, $E=\mpl\,{\rh}/{2\,\lp}$, and define the
horizon wave-function
\be
\psih(\rh)
=
{\mathcal{N}_{\rm H}}\,C(\rh(E))
\ ,
\ee
whose normalisation $\mathcal{N}_{\rm H}$ can be fixed by using the norm
defined by the scalar product
\be
\pro{\psih}{\phi_{\rm H}}
=
4\,\pi
\int_0^\infty
\psih^*(\rh)\,\phi_{\rm H}(\rh)\,\rh^2\,\d \rh
\ .
\label{prod}
\ee
Let us remark again that this quantum description of the gravitational radius can be
viewed as a particular way of quantising the Einstein equation~\eqref{einstein},
which lifts to the quantum level the relation between $\Rh$ and the Misner-Sharp mass
$m$ determined by the matter source according to Eq.~\eqref{M}.
In other words, we are assuming that the only relevant quantum degrees of freedom
associated with the gravitational structure of space-time (which classically give rise to
trapping surfaces) are those determined by the quantum degrees of freedom of
the matter source.
This implies that, at least in the static case, we can just consider ``on-shell'' states
for the gravitational degrees of freedom [the ``mass-shell'' relation being here precisely
Eq.~\eqref{einstein} viewed as an operator equation], and neglect the contribution of
``purely gravitational'' fluctuations.
The latter could then be studied by employing standard background field method
techniques, in the same way one determines the Lamb shift for the hydrogen atom.
\par
We can interpret the normalised wave-function $\psih$ as yielding the probability
that we would detect a gravitational radius of areal radius $r=\rh$ associated
with the particle in the quantum state $\psis$.
Such a radius is necessarily ``fuzzy'', like are the energy and the position of the
particle itself, and will have an uncertainty 
\be
\Delta\rh 
=
\sqrt{\expec{\rh^2}-\expec{\rh}^2}
\ .
\ee
Moreover, having defined the $\psih$ associated with a given $\psis$,
we can also define the conditional probability density that the particle lies
inside its own gravitational radius $\rh$ as
\be
\mathcal{P}_<(r<\rh)
=
P_{\rm S}(r<\rh)\,\mathcal{P}_{\rm H}(\rh)
\ ,
\label{PrlessH}
\ee
where
\be
P_{\rm S}(r<\rh)
=
4\,\pi\,\int_0^{\rh}
|\psis(r)|^2\,r^2\,\d r
\ee
is the usual probability that the particle is found inside a sphere of radius $r=\rh$,
and
\be
\mathcal{P}_{\rm H}(\rh)
=
4\,\pi\,\rh^2\,|\psih(\rh)|^2
\ee
is the probability density that the gravitational radius is located on the sphere of radius $r=\rh$.
Since this is the analogue of the condition~\eqref{Ehor}, one can also view $\mathcal{P}_<(r<\rh)$
as the probability density that the sphere $r=\rh$ is a trapping surface.
Finally, the probability that the particle described by the wave-function $\psis$ is a
black hole (regardless of the horizon size), will be obtained by integrating~\eqref{PrlessH}
over all possible values of $\rh$, namely
\be
P_{\rm BH}
=
\int_0^\infty
\mathcal{P}_<(r<\rh)\,\d \rh
\ ,
\ee
which will depend on the observables and parameters of the specific state $\psis$ we started
the construction.
Let us again emphasise that we have frozen any possible time dependence 
so far, or, equivalently, the above quantities are only defined at a given instant of time.
\subsection{Single massive particle}
As the simplest example, we shall consider a massive particle at rest
described by the Gaussian wave-function~\cite{fuzzyh,gupf}
\be
\psis(r)
=
\frac{e^{-\frac{r^2}{2\,\ell^2}}}{\ell^{3/2}\,\pi^{3/4}}
\ .
\label{Gr}
\ee
From the usual flat-space normal modes
\be
j_0(p\,r)
=
\frac{\sin(p\,r)}{\sqrt{2^3\,\pi^3}\,p\,r}
\ ,
\ee
one obtains the momentum space wave-function
\be
\psis(p)
\equiv
4\,\pi\,\int_0^\infty
\psis(r)\,j_0(p\,r)\,r^2\,\d r
=
\frac{e^{-\frac{p^2}{2\,\Delta^2}}}{\Delta^{3/2}\,\pi^{3/4}}\,
\ ,
\label{Gp}
\ee
where $p^2=\vec p\cdot\vec p$ and $\Delta=\mpl\,\lp/\ell$.
For the energy of the particle, in accord with the above choice of modes,
we simply assume the relativistic mass-shell relation
in flat space,
\be
E^2
=
p^2+m^2
\ge
m^2
\ ,
\label{freeH}
\ee
and we treat $\lambda_m \simeq \lp\,{\mpl}/{m}$ and $\ell$ as two
independent length scales, although it is reasonable to assume that
\be
\ell\gtrsim\lambda_m
\ ,
\ee
since one does not expect it is physically possible to locate a particle
with an accuracy greater than its Compton length.
Upon expressing $E$ in terms of the Schwarzschild radius,
$E=\mpl\,{\rh}/{2\,\lp}$, we obtain the horizon wave-function
\be
\psih(\rh)
\simeq
\Theta(\rh-r_{\rm m})\,
\exp\left[
-\frac{\ell^2}{8\,\lp^4}\,
\left(\rh^2-r_{m}^2\right)
\right]
\ ,
\label{Gh}
\ee
where
\be
r_{m}
=
2\,\lp\,\frac{m}{\mpl}
\ee
is the minimum allowed value of $\rh$ corresponding to the
minimum value of $E=m$, and $\Theta$ is Heaviside's step function.
The normalisation of $\psih$ should then be fixed in the scalar
product~\eqref{prod}.
Since, for the minimum width $\ell\simeq\lambda_m$, one has
$\expec{\rh}\simeq \lp^2/\ell\simeq \Rh$ and  
\be
\Delta\rh 
=
\sqrt{\expec{\rh^2}-\expec{\rh}^2}
\simeq
{\lp^2}/{\ell}
\ .
\ee
one finds the uncertainty $\Delta\rh \simeq \Rh$,
which is dangerously large for a macroscopic black hole with $\Rh\gg\lp$.
In fact, this result could be viewed as supporting the picture that 
macroscopic black holes should be made of a very large number $N$ of very
light particles, precisely like the BEC of gravitons of Refs.~\cite{dvali},
for which one instead obtains $\Delta\rh/\expec{\rh}\sim 1/N$~\cite{BEC_BH}.
\par
\begin{figure}[t]
\centering
\raisebox{4cm}{$P_{\rm BH}$}
\includegraphics[width=6.5cm,height=5cm]{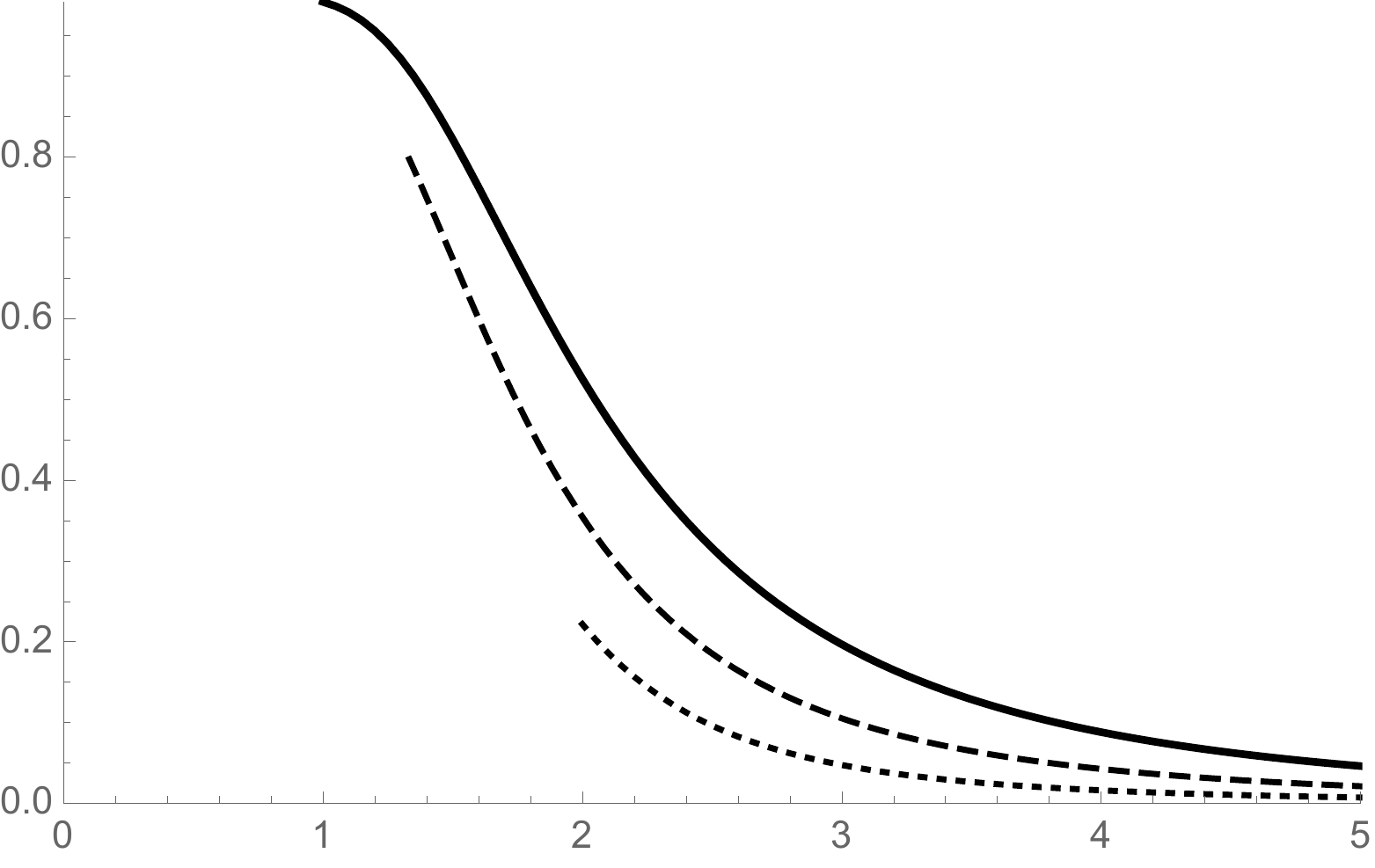}
$\ell/\lp$
$\quad$
\includegraphics[width=6.5cm,height=5cm]{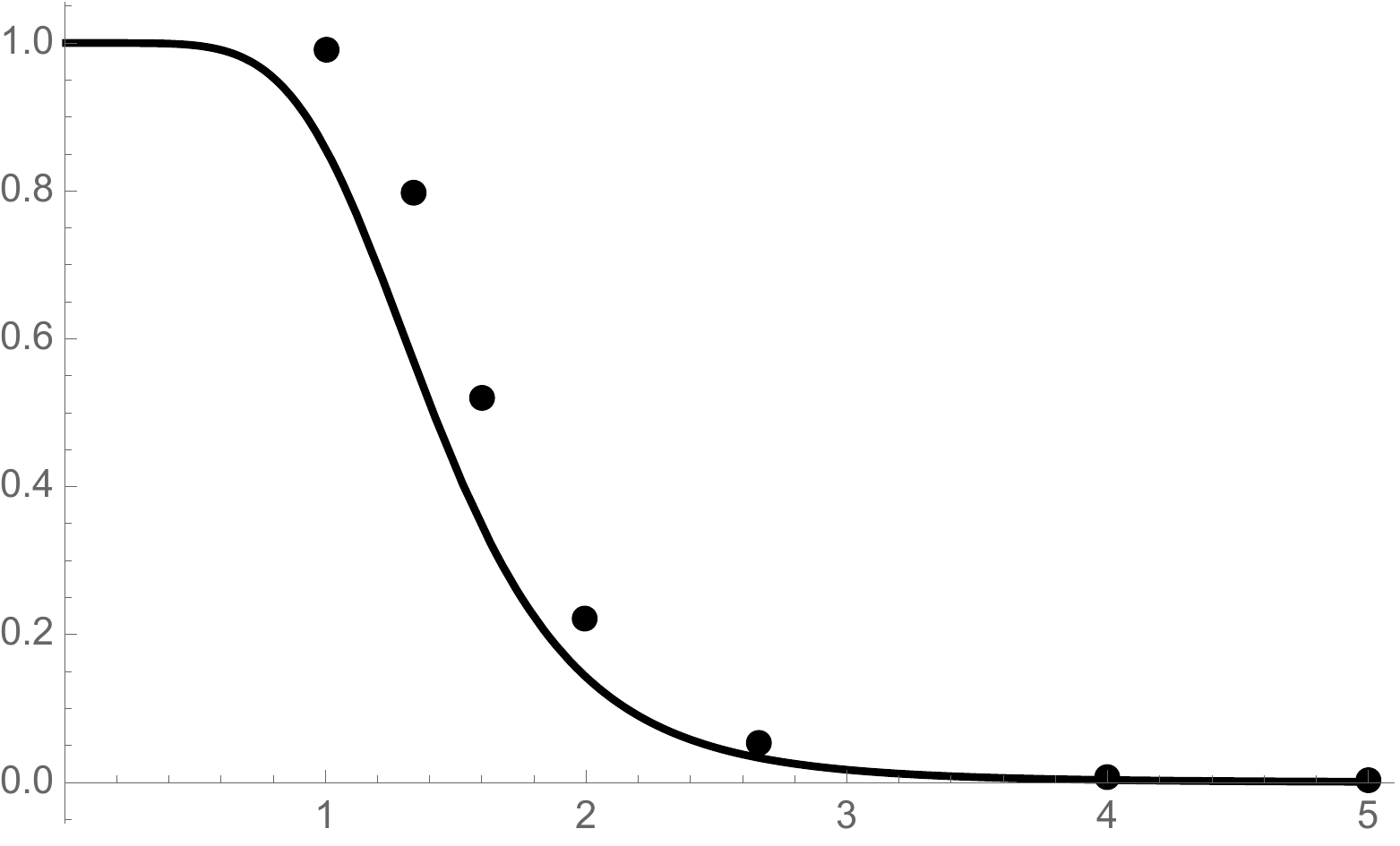}
$\ell/\lp$
\caption{Left panel: probability that a particle of 
Gaussian width $\ell\ge\lambda_m$ is a black hole for mass
$m=\mpl$ (solid line), $m=3\,\mpl/4$ (dashed line)
and $m=\mpl/2$ (dotted line).
Right panel: probability that a particle of 
Gaussian width $\ell=\lambda_m$ is a black hole (dots)
compared to its analytical approximation~\eqref{Pbha}.
\label{prob}}
\end{figure}
From $\psis$ in Eq.~\eqref{Gr} and the normalised $\psih$ from Eq.~\eqref{Gh},
one can numerically obtain the probability $P_{\rm BH}=P_{\rm BH}(\ell;m)$,
which is plotted in the left panel of Fig.~\ref{prob}, for a few values of
$m$ about and below the Planck scale, and $\ell\ge \lambda_m$.
On considering the limiting case $\ell=\lambda_m$, one obtains
$P_{\rm BH}=P_{\rm BH}(\ell)$ displayed by the dots in the right panel
of Fig.~\ref{prob}.
It is then interesting to notice that these points are fairly well approximated
by the analytical expression used in Refs.~\cite{fuzzyh,gupf}.
By formally taking the limit $r_{m}\to 0$ in Eq.~\eqref{Gh}
and normalising according to Eq.~\eqref{prod}, one obtains
\be
\psih(\rh)
=
\frac{\ell^{3/2}\,e^{-\frac{\ell^2\,\rh^2}{8\,\lp^4}}}
{2\,\sqrt{2}\,\lp^{3/2}\,\pi^{3/4}}
\ ,
\label{Gha}
\ee
which then leads to 
\be
P_{\rm BH}(\ell)
=
\frac{2}{\pi}\left[
\arctan\left(2\,\frac{\lp^2}{\ell^2}\right)
-
\frac{2\,\ell^2\,(\ell^4/\lp^4-4)}{\lp^2\,(4+\ell^4/\lp^4)}
\right]
\ .
\label{Pbha}
\ee
This probability is represented by the solid line in the right panel 
of Fig.~\ref{prob}, where one can see that it only underestimates
the correct probability for values of $m\gtrsim\mpl/2$
(that is, $\ell\lesssim 2\,\lp$).
\section{Causal time evolution}
\label{Tevo}
\setcounter{equation}{0}
Let us now try and investigate the time evolution of the simple toy model
of a spherically symmetric particle described above.
In order to disentangle the effects on the dynamics due to the horizon from other
physics, we shall neglect any non-gravitational interaction, and assume that the UV
completion of gravity at the Planck scale does not introduce any new
mechanisms~\footnote{Let us further point out the former simplification would clearly
be unphysical for standard model particles, whereas the latter is at least debatable.
The resulting dynamical picture will correspondingly be simplistic, but it will allow us
to carry out a complete analysis of the Gaussian particle.}. 
\par
From the point of view of an observer located sufficiently far from the particle,
we can assume we have a reliable knowledge of two limiting cases:
\begin{description}
\item[ a) ]
if the particle is not a black hole ($P_{\rm BH}\ll 1$), the evolution
occurs according to the usual laws of quantum mechanics.
For the free massive particle described by the Gaussian packet~\eqref{Gr},
this means free evolution according to the Hamiltonian $H=E$ in
Eq.~\eqref{freeH};
\item[ b) ]
if the system is a black hole ($P_{\rm BH}\simeq 1$), no evolution appears
to occur at all.
Note we are specifically neglecting Hawking evaporation in this simplified
hypothesis, and view a black hole like a ``frozen star''~\footnote{This was the name
most commonly used for gravitationally collapsed objects before the term black hole
was introduced in the sixties.}.
\end{description}
When the particle is in a generic quantum state $\psis$ not satisfying any
of the above two limiting conditions, we then assume the evolution for ``short''
time intervals $\delta t$ is ruled by the simple prescription
\be
\psis(r,t+\delta t)
=
\left[
\mu_{\rm H}(t)\,\hat{\mathbb{I}}
+
\bar\mu_{\rm H}(t)\,e^{-\frac{i}{\hbar}\,\hat H\,\delta t}
\right]
\psis(r,t)
\ ,
\label{dpsi}
\ee
where $\hat{\mathbb{I}}$ is the identity operator, $\mu_{\rm H}(t)\simeq P_{\rm BH}(t)$
and $\bar\mu_{\rm H}(t)\simeq 1-P_{\rm BH}(t)$, so that the two limiting behaviours a) and b)
are both accounted for by construction.
In fact, we first note that, upon assuming $\mu_{\rm H}$ and $\bar\mu_{\rm H}$ are real,
unitarity is preserved provided these coefficients satisfy the normalisation condition
\be
1
&\!\!=\!\!&
\mu_{\rm H}^2+\bar\mu_{\rm H}^2
+2\,\bar\mu_{\rm H}\,\mu_{\rm H}\,\cos\left((\delta t/\hbar)\,\hat H\right)
\nonumber
\\
&\!\!=\!\!&
\left(\mu_{\rm H}+\bar\mu_{\rm H}\right)^2
+\mathcal{O}(\delta t^2)
\ ,
\label{muNre}
\ee
or $\bar\mu_{\rm H}\simeq 1-\mu_{\rm H}$.
The evolution equation then becomes
\be
i\,\hbar\,\frac{\delta \psis(r,t)}{\delta t}
\simeq
\left[1-P_{\rm BH}(t)\right]
\hat H\,
\psis(r,t)
\ ,
\label{schro}
\ee
which has the form of an effective Schr\"odinger equation, and
the standard quantum mechanical evolution is exactly recovered in the limit
$P_{\rm BH}\to 0$ [limiting case~a)].
Note, however, that $P_{\rm BH}=P_{\rm BH}(t)$ depends on the whole wave-function
$\psis=\psis(r,t)$ and the apparently trivial correction it entails is actually highly
non-local, in the sense that it cannot be straightforwardly reproduced by an interacting
term of the form $H_{\rm int}=H_{\rm int}(r,t)$.
\par
This observation makes it clear that it will in general be very difficult to solve
the above equation~\eqref{schro} for a finite time interval.
It is instead easier to stick with Eq.~\eqref{dpsi},
and employ the spectral decomposition at time $t$,
\be
\psis(r,t)
=
\sum_E
C_E(t)\,j_0(E,r)
\ ,
\label{CEt}
\ee
which then leads to
\be
i\,\hbar\,\delta C_E(t)
\simeq
\left[1-P_{\rm BH}(t)\right]
E\,
C_E(t)\,
{\delta t}
\ ,
\label{CETeq}
\ee
where we recall the time-dependent coefficient $P_{\rm BH}=P_{\rm BH}(t)$
is determined by the whole wave-function $\psis=\psis(r,t)$.
From Eq.~\eqref{CETeq}, and knowing $\psis$ at the time $t$,
one can reconstruct both $\psis$ and $\psih$ at $t+\delta t$.
These two wave-functions will then allow one to proceed to the next time step. 
We also notice that the approximation~\eqref{muNre} requires
\be
\delta t
\lesssim
\frac{\hbar}{E}
=
\lp\,\frac{\mpl}{E}
\ ,
\label{dt}
\ee
which shall be duly commented on, and properly taken into account, in the following.
\subsection{Short time evolution}
\label{short}
The above evolution equations can be applied to the Gaussian wave-function~\eqref{Gr},
for which the initial probability $P_{\rm BH}(t=0)=P_{\rm BH}(\ell;m)$ was computed
numerically in the previous Section. 
From that result, we expect the particle will likely be a black hole only if $m\gtrsim\mpl$
and $\ell\lesssim\lp$.
\par
In particular, according to Eq.~\eqref{dt} with $m\simeq \mpl$, we expect our evolution
equation~\eqref{CETeq} holds for
\be
\delta t\lesssim \hbar/E\simeq \lp
\ ,
\label{dtp}
\ee
and shorter for modes with energy $E>\mpl$.
One might question the validity of our approach for trans-Planckian modes,
and such sub-Planckian times.
First of all, it is interesting to note that the duality $(E>\mpl)\Leftrightarrow (\delta t<\lp)$ 
clearly appears in this dynamics.
Further, we have already commented that the evolution equation~\eqref{dpsi}
could only hold in the (admittedly unrealistic approximation) that non-gravitational
forces can be neglected and the UV completion of gravity is trivial.
Different cases could be considered by modifying the dispersion
relation~\eqref{freeH}, but we shall see that our very strong simplifications
nonetheless allow us to draw some interesting (qualitative) conclusions.
\par
\begin{figure}[t]
\centering
\raisebox{4cm}{${\mathcal P}_{\rm S}$}
\includegraphics[width=10cm,height=5cm]{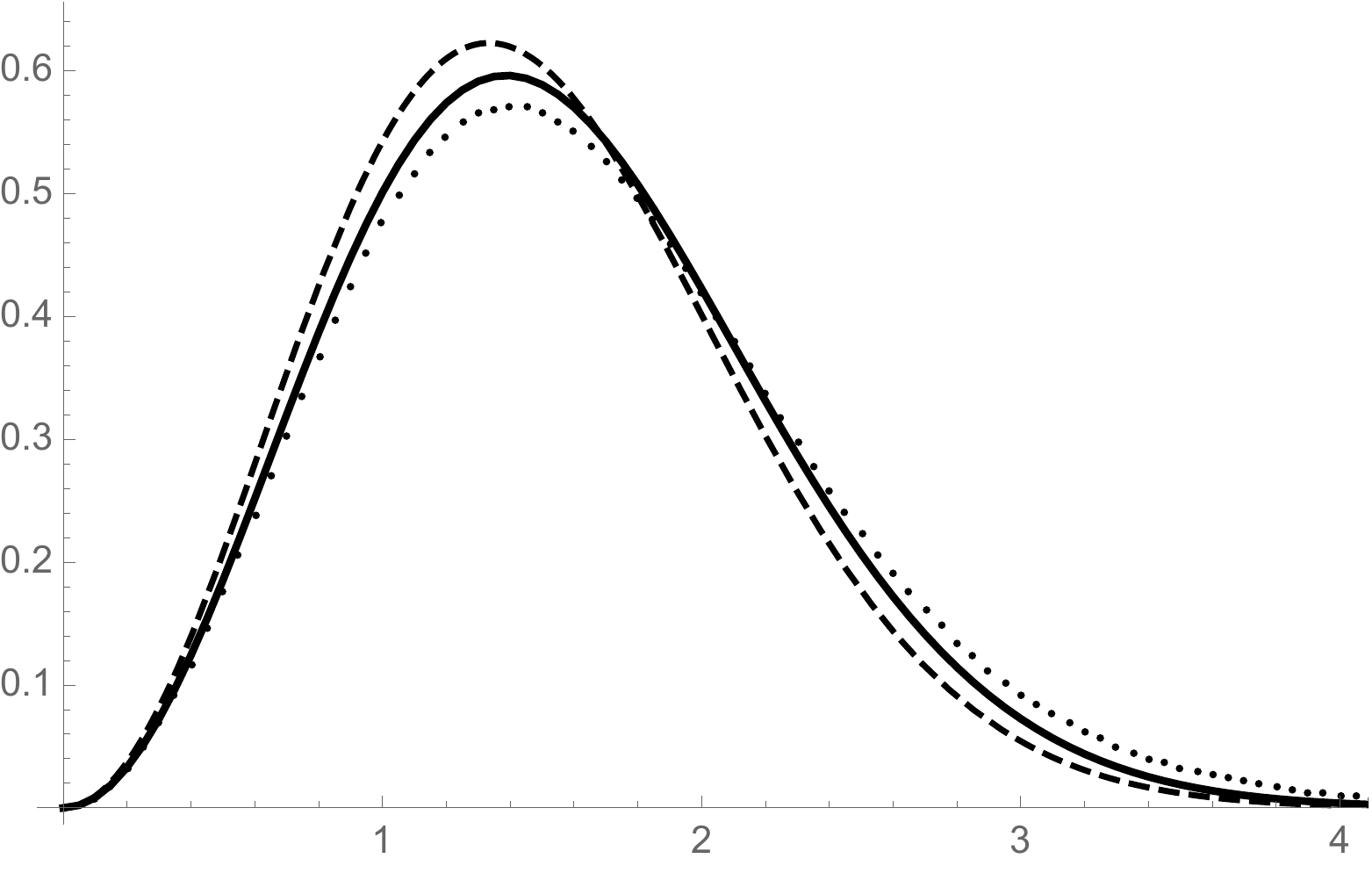}
\raisebox{0cm}{$\frac{r}{\lp}$}
\caption{Time-evolution of the probability density for the initial Gaussian packet~\eqref{Gr}
with $m=3\,\mpl/4$ and $\ell=\lambda_m=4\,\lp/3$ (dashed line) according to standard
quantum mechanics (dotted line) compared to its causal evolution~\eqref{schro} (solid line)
for $\delta t=\lp$.
\label{T1_2}}
\end{figure}

We now proceed to solve Eq.~\eqref{CETeq} with the time step~\eqref{dtp},
and then invert the spectral decomposition~\eqref{CEt} in order to reconstruct
the wave-function $\psis$ at the time $t=\delta t$.
This procedure can be carried out analytically, provided the probability $P_{\rm BH}(\ell;m)$
is given, but the result is better described graphically.
In Figs.~\ref{T1_2}, we plot the probability density ${\mathcal P}_{\rm S}=4\,\pi\,r^2\,|\psis(r,t)|^2$
at $t=0$ and $t=\delta t=\lp$ obtained from the modified evolution~\eqref{schro}, for 
a value of the mass of $m=3\,\mpl/4$ and $\ell=\lambda_m=4\,\lp/3$,
and we confront it with the probability density obtained from the standard free evolution
at the same time $t=\lp$. 
This case corresponds to a minimum gravitational radius $r_m=1.5\,\lp$, an expectation vale
of the energy $\expec{E}\simeq 1.15\,\mpl$, a Schwarzschild radius
$\Rh=\expec{\hat r_{\rm H}}\simeq 2.3\,\lp$,
and initial probability $P_{\rm BH}\simeq 0.8$.
The time step $\delta t=\lp$ saturates (and likely exceeds) the allowed bound, but was
chosen in order to make it clear that the packet remains more confined than one would
expect according to the usual quantum mechanical evolution.
Since the packet still spreads, one expects the probability $P_{\rm BH}(t+\delta t)<P_{\rm BH}(t)$,
and the effect of the horizon will weaken over time.
In order to support this expectation, we need to be able to study the time evolution over
more then one time step.
\subsection{Long time evolution}
\label{long}
\begin{figure}[t!]
\centering
\raisebox{4cm}{${\mathcal P}_{\rm S}$}
\includegraphics[width=7cm,height=5cm]{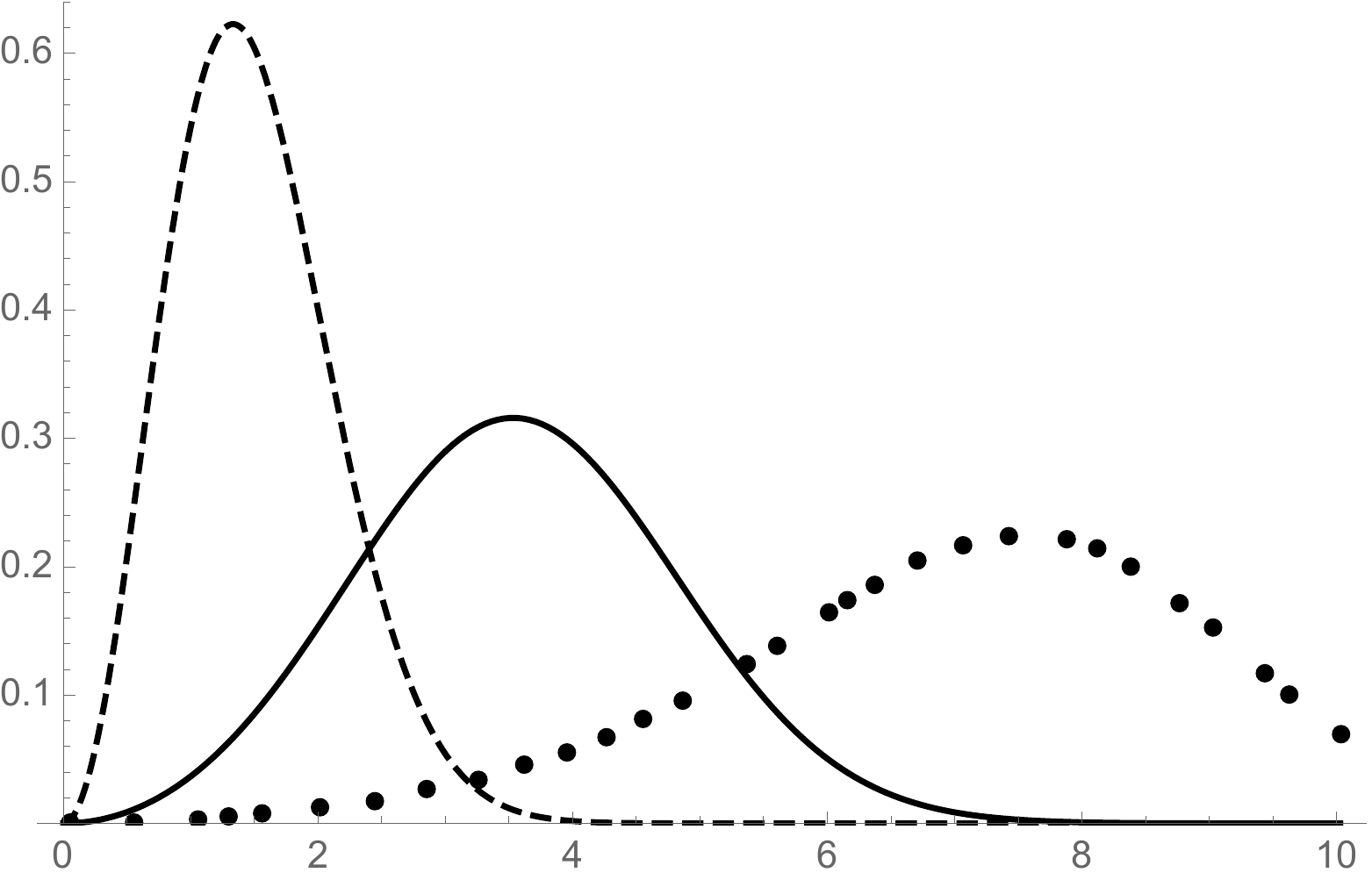}
\raisebox{0cm}{$\frac{r}{\lp}$}
\raisebox{4cm}{${\mathcal P}_{\rm H}$}
\includegraphics[width=7cm,height=5cm]{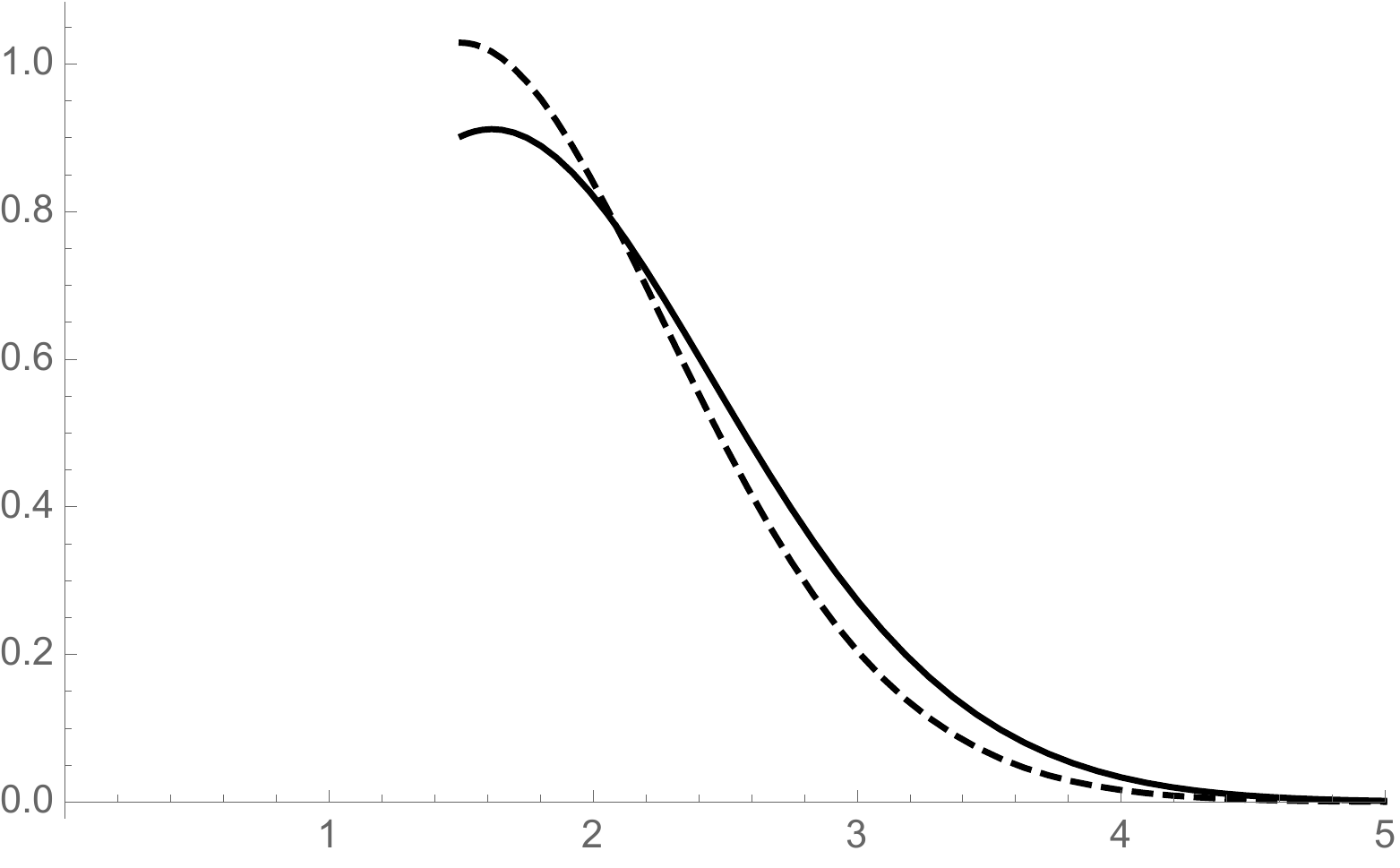}
\raisebox{0cm}{$\frac{\rh}{\lp}$}
\caption{Left panel: probability density from the final wave-packet $\psis^{\rm fin}=\psis(r,10\,\lp)$
with $\ell=\lambda_m$ and $m=3\,\mpl/4$ obtained from the modified evolution~\eqref{schro}
(solid line) compared to the freely evolved packet (dotted line) and initial packet 
$\psis^{\rm in}=\psis(r,0)$ (dashed line). 
Right panel: horizon probability density for the Gaussian particle in the left panel at $t=0$
(dotted line) and $t=10\,\lp$ (solid line).
Note that $\psih(\rh <r_m,t)=0$, for $r_m=1.5\,\mpl$.
\label{Ppsit}}
\end{figure}
\begin{figure}[t]
\centering
\raisebox{4cm}{$P_{\rm BH}$}
\includegraphics[width=8cm,height=5cm]{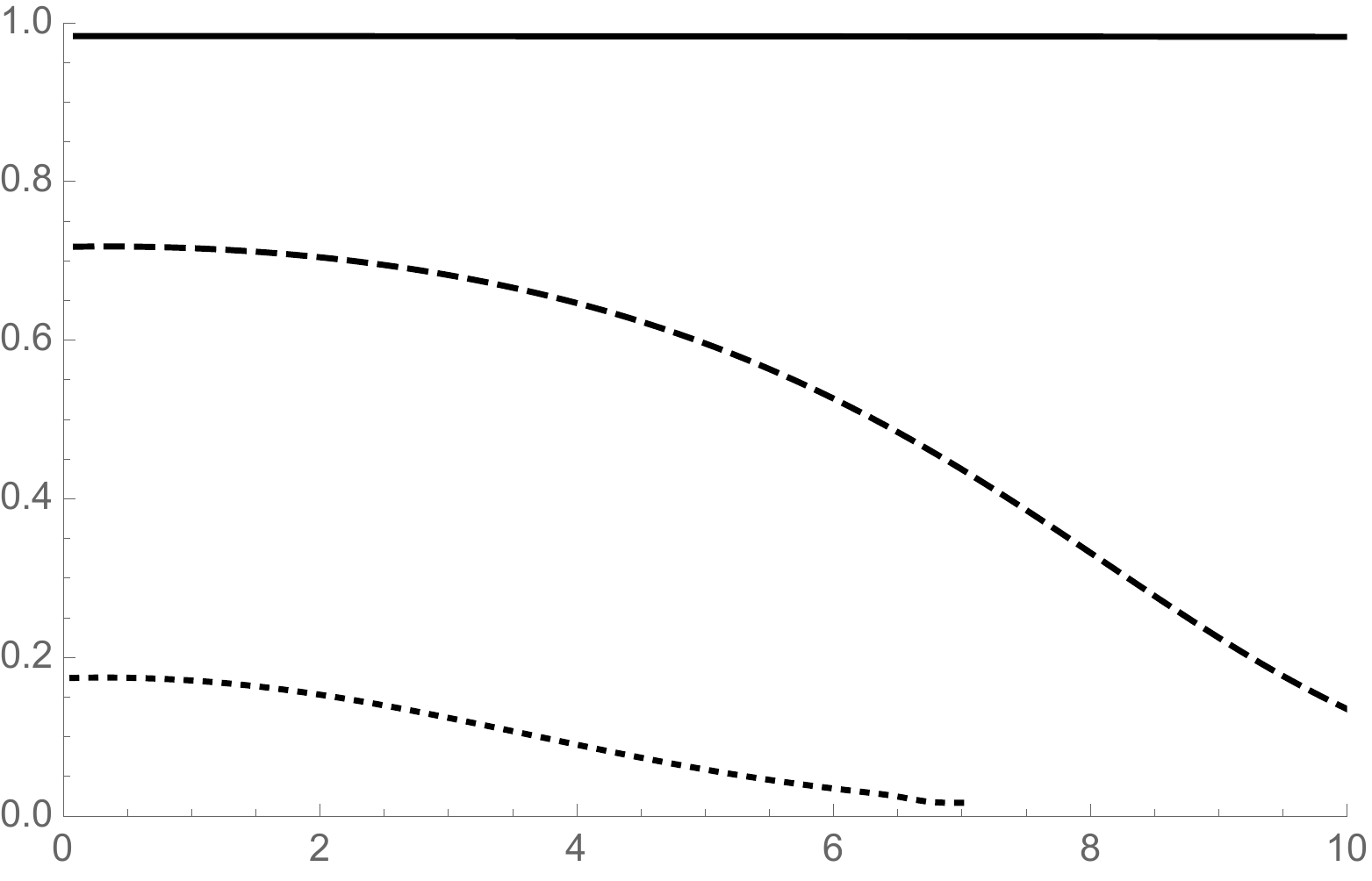}
\raisebox{0cm}{$\frac{t}{\lp}$}
\caption{Time-evolution of the probability $P_{\rm BH}$ for the Gaussian
wave-function~\eqref{Gr} with $\ell=\lambda_m$ for $m=\mpl$ (solid line)
$m=3\,\mpl/4$ (dashed line) and $m=\mpl/2$ (dotted line).
\label{Pbht}}
\end{figure}
In order to study the time evolution for longer times, we discretise the whole interval of
time by writing $t=n\,\delta t$, where $1\le n<N$ is a positive integer, and we recall the time
step $\delta t$ must satisfy the bound~\eqref{dt} for all relevant energies $E$ in the spectral
decomposition~\eqref{CEt} and at all steps $n$ (this will be kept under control numerically).
We can then iteratively compute $P_{\rm BH}(n\,\delta t)$ in order to determine
$\psis(r,(n+1)\,\delta t)$ by solving Eq.~\eqref{schro}, or equivalently Eq.~\eqref{CETeq},
and from that compute $P_{\rm BH}((n+1)\,\delta t)$.
This procedure can be implemented numerically in order to compute the
final state $\psis^{\rm fin}=\psis(r,N\,\delta t)$ starting from any given initial state
$\psis^{\rm in}=\psis(r,0)$.
\par
The time evolution of the probability density ${\mathcal P}_{\rm S}$
for the same wave-packet $\psis$ with $m=3\,\mpl/4$ and
$\ell=\lambda_m=4\,\lp/3$ shown in Fig.~\ref{T1_2}, is displayed in the left panel of
Fig.~\ref{Ppsit}, now at the final time $t=10\,\lp$.
From this plot one can see that the packet indeed spreads more slowly
than predicted by the standard free evolution, but it still gets wider and wider,
so that one expects the probability for the system to be a black hole will decrease.
The right panel of Fig.~\ref{Ppsit} also shows that the corresponding initial and final
horizon probability densities ${\mathcal P}_{\rm H}=4\,\pi\,{\rh}^2\,\psih^2$ do not differ
very significantly, with the peaks located roughly around the same value of $\rh$,
in agreement with energy conservation.
It is in fact more interesting to display directly the time evolution of the probability
$P_{\rm BH}=P_{\rm BH}(t)$.
The latter can be seen in Fig.~\ref{Pbht}, for $\ell=\lambda_m$,
and with the same values of $m=\mpl$, $m=3\,\mpl/4$ and $m=\mpl/2$ used to produce
Fig.~\ref{prob}.
It is clear that for masses $m<\mpl$, the probability that the system remains a black hole
drops very quickly in time.
This could be interpreted as a decay of the initial (``fuzzy'') black hole, ``without'' the
well-known Hawking radiation at work (or, alternatively, with the Hawking radiation
mimicked by the spreading of the packet). 
Let us however notice that, even putting all of our strong simplifications aside,
the results presented here would only apply to particle-like toy 
objects of extremely large masses $m\sim\mpl$ and with a minimum energy
$E\simeq m\sim\mpl$, whose existence is not known in nature.
Composite objects, like the more realistic BEC models of black holes of Refs.~\cite{dvali},
will have to be analysed separately, and the Hawking radiation they naturally describe
will have then to be explicitly accommodated for in the dynamical picture.
\section{Conclusions}
\label{conc}
\setcounter{equation}{0}
In this work, we have extended the proposal of a ``horizon wave-function''
$\psih$ from Refs.~\cite{fuzzyh} by showing how it can be also used to describe the
time evolution of a quantum system with a non-negligible probability to
contain a ``fuzzy'' trapping surface.
We have done so by assuming the evolution operator for a black hole is
exactly the identity, which means no evolution would occur for an ideal state
with probability $P_{\rm BH}=1$ (the ``frozen star'' picture raised to the quantum level).
For a general state, the time evolution is instead obtained by
replacing $\hat H\to (1-P_{\rm BH})\,\hat H$ in the Schr\"odinger equation.
This replacement clearly preserves unitarity and also energy conservation
(for isolated systems), although the coefficient $P_{\rm BH}=P_{\rm BH}(t)$
depends on the whole wave-function $\psis=\psis(r,t)$ [and the corresponding
$\psih=\psih(r,t)$], and cannot therefore be simply derived from a local interaction
term.
One could further speculate that the proper quantum evolution operator for a black
hole should describe the Hawking evaporation, thus differing from the simple
identity we assumed here.
As we mentioned at the end of the previous Section, this aspect can likely be
better addressed when considering more realistic models of black holes than
the toy Gaussian particle we used here. 
\par
At the formal level, one could argue the above picture should be substantiated
by providing a more detailed construction of the Hilbert spaces of the two
wave-functions $\psis$ and $\psih$, and thoroughly analysing the properties of
the operator that maps the former into the latter.
This construction should indeed be possible by starting explicitly from the canonical
quantisation of the relevant Einstein equation~\eqref{einstein}, albeit most likely
in a formalism that does not make use of the Arnowitt-Deser-Misner decomposition
of space-time that leads to the the usual Wheeler-DeWitt equation.
It might also be interesting to develop an alternative (possibly equivalent)
prescription for the time evolution using a functional integral formalism,
with different paths weighted by $\psih$ or $P_{\rm BH}$.
Finally, it will certainly be important to generalise the whole approach to
sources described by quantum field theory, in order to take into account known
properties of standard model particles and interactions, as well as Lorentz invariance.
In particular, our approach appears so far non-relativistic, in that it was always
applied in the centre-of-mass reference frame in which the (forming or
decaying) black hole is at rest~\cite{acmo}, and the Planck scale is
therein naturally related with the proper mass and radius of a particle~\cite{Clor}. 
These very important aspects are left for future investigations, but we would like to
end by noting the key role played by localisation in this business~\cite{Thorne:1972ji}.
In order to understand black hole formation and decay, it is unlikely
one can neglect the spatial configuration of the physical system and
just work in momentum space.
This observation immediately brings to mind long-standing issues,
like the very possibility of defining the position of a particle in quantum field
theory~\cite{NW}.
\section*{Acknowledments}
It is a pleasure to thank A.~Orlandi, F.~K\"uhnel and O.~Micu 
for stimulating discussions and comments.
\end{document}